\documentclass[lettersize,journal]{IEEEtran}
\usepackage{amsmath,amsfonts}
\usepackage{algorithmic}
\usepackage{algorithm}
\usepackage{array}
\usepackage[caption=false,font=normalsize,labelfont=sf,textfont=sf]{subfig}
\usepackage{textcomp}
\usepackage{stfloats}
\usepackage{url}
\usepackage{verbatim}
\usepackage{graphicx}
\usepackage{cite}
\usepackage{booktabs}
\usepackage{hyperref}
\usepackage{orcidlink}
\usepackage{multirow}

\usepackage{tikz}
\usetikzlibrary{positioning, fit, backgrounds, arrows.meta, calc}

\begin{document}

\title{GhostFaceMPC: Privacy-Preserving Face Recognition with Multi-Party Computation}

\author{Davide Carnemolla~\orcidlink{0009-0001-2575-0874}, Dario Catalano~\orcidlink{0000-0001-9677-944X}, Mario Di Raimondo~\orcidlink{0000-0002-5531-1098}

\thanks{This work was funded in parts by the PIAno di in-
CEntivi per la Ricerca di Ateneo 2024/2026, linea di intervento 1, University of Catania.}
\thanks{Davide Carnemolla, Dario Catalano, and Mario Di Raimondo are with the Department of Mathematics and Computer Science, University of Catania, Italy (e-mail: davide.carnemolla@phd.unict.it; dario.catalano@unict.it; mario.diraimondo@unict.it).}
}



\maketitle

\begin{abstract}
Facial recognition has become central to modern authentication systems, security frameworks, and personalized services. However, its widespread adoption is hindered by severe privacy risks, particularly surrounding unauthorized access to sensitive biometric data. To address these challenges, this paper presents GhostFaceMPC, the first end-to-end, privacy-preserving face recognition system to evaluate a deep convolutional neural network extractor and matching under secure multi-party computation. GhostFaceMPC enables secure processing of facial data across all stages of a face recognition process, from feature extraction and storage to matching, without exposing raw images or biometric embeddings at any stage.
We introduce a multi-party computation-friendly quantization of the GhostFaceNet model and circuit optimizations that make face recognition under multi-party computation practical. The code is available at \url{https://github.com/crypto-unict/GhostFaceMPC}.
\end{abstract}


\section{Introduction}
\label{sec:introduction}
\IEEEPARstart{F}{ace} Recognition (FR)~\cite{arcface, AdaFace, CosFace} systems have become integral to modern identity management, spanning applications from personal device authentication to law enforcement and financial services. However, these systems rely on highly sensitive biometric data whose exposure may induce privacy violations, identity theft, and unauthorized surveillance. Unlike traditional credentials, biometric traits are inherently immutable: once compromised, they cannot be revoked or reset. This permanence underscores the critical need for robust, privacy-preserving mechanisms to safeguard biometric data.

Such protections are also enforced by legal frameworks that govern the acquisition, storage and processing of biometric data~\cite{PE-FB-survey}, \emph{e.g.}, the General Data Protection Regulation (GDPR) of the European Union~\cite{Voigt2024}. In practical deployments, a standard FR pipeline comprises three main components: a probe face image, a feature extractor (typically a deep neural network), and a reference database of biometric templates. The feature extractor processes an input image to yield a compact, discriminative feature embedding.

The embedding is then used either for verification (1:1), in which the probe is compared against the template of a claimed identity, or for identification (1:N), in which it is compared against every enrolled template. GhostFaceMPC supports both tasks.

\subsection{Contributions of our work}
Most cryptographic face recognition systems protect only the matching stage and
leave feature extraction to the client (see Section~\ref{sec:related}). A nice exception 
is CryptoFace~\cite{cryptoface} that relies on FHE \cite{STOC:Gentry09} and, to make the usage of FHE practical, it forces a shallow
patch-based backbone which induces an overhead of over 20 minutes per image. 
In this paper, we show that a
standard deep face network can be evaluated end to end under MPC at practical
costs and with negligible accuracy loss. Our contributions can be summarized as follows: 
\begin{itemize}
\item \textbf{End-to-end protected pipeline.} We present GhostFaceMPC,
the first (to the best of our knowledge) face recognition system that evaluates a deep Convolutional Neural Network (CNN) extractor,
matching and thresholding under MPC. The client splits the (aligned) face image into random shares sent to the servers. The three servers 
(jointly) compute the feature, compare it with the enrolled template of the claimed identity and apply the threshold, all on shares. 
Finally, the servers return shares of a single bit (match or no match) that only the client reconstructs. 

No single server learns the image, the feature vector, the enrolled templates or the result. Moreover, the model is known in the clear only to the servers and is never disclosed to the client, which is valuable in deployments where the servers run a proprietary model. GhostFaceMPC supports semi-honest security and
security with abort against one malicious server (see Section~\ref{sec:arch-and-threat}).

\item \textbf{MPC-friendly quantization.} Since the communication of every
interactive operation grows with the width of the underlying ring, we evaluate the network over
$\mathbb{Z}_{2^{32}}$ instead of $\mathbb{Z}_{2^{64}}$. However, naively encoding the
trained floating-point model at this precision loses over 30 percentage points
of accuracy, because rounding errors accumulate across layers. We therefore
fine-tune a copy of the model (the student) to reproduce the embeddings of the
original (the teacher), with a forward pass that simulates the fixed-point
encoding and roundings executed by the servers. The resulting model halves the
communication per element and stays within 0.32 percentage points of the
original accuracy on five benchmarks (Section~V-B).

\item \textbf{Circuit optimizations.} 
Since the weights are public, every product of a secret activation by a weight is local. The interactive operations are the fixed-point roundings after each linear output, the sign extractions of the activation functions, and a few secret-by-secret products, with roundings and sign extractions dominating the communication.
We reduce both the number and the cost of these roundings. We merge each batch
normalization into the preceding convolution. Wherever a bound calibrated on held-out faces indicates that the ring does not overflow, we postpone roundings across consecutive layers so
that one rounding replaces two. We combine known techniques for replicated sharing~\cite{ABY3,dalskov2020} into an exact two-summand rounding that halves the bit conversions per carry, and we
vectorize each layer so that its roundings share the same communication rounds.
Together, these reduce online communication by 70\%, rounds by 96\% and latency
by 95\% relative to a direct MP-SPDZ implementation (Section~\ref{subsec:optimizations}).

\item \textbf{Implementation and 
Evaluation.} We implemented GhostFaceMPC in
MP-SPDZ~\cite{mp-spdz} and evaluated it on five standard benchmarks. In the semi-honest setting, a verification request (one probe against the template of the claimed
identity) takes 2.4\,s and 312\,MB of online communication when the three
servers run on a single machine, and 3.8\,s on a LAN. When considering  security with aborts, the online phase takes 5.5\,s on a single machine. 
Notice that executing the non-quantized model under
the general-purpose EzPC framework (Porthos protocol) takes 259\,s and 18\,GB.
GhostFaceMPC also supports closed-set identification, in which the probe is
compared against all enrolled identities (Section~\ref{sec:experiments}). The code and the
quantized model are publicly available~\footnote{\url{https://github.com/crypto-unict/GhostFaceMPC}}.
\end{itemize}

\subsection{Organization of this paper}
The paper is structured as follows. Section~\ref{sec:related} deals with related work. Section~\ref{sec:preliminaries} introduces the necessary preliminaries. Section~\ref{sec:arch-and-threat} details the system architecture and defines the threat model. In Section~\ref{sec:ghostfacempc} we present our solution, highlighting system design choices, implementation details, and optimizations. Section~\ref{sec:experiments} presents the experimental evaluation and results.
Finally, in Section~\ref{sec:conclustion} we draw our conclusions.

\section{Related Work}
\label{sec:related}
Privacy-preserving face recognition (PPFR) has been pursued with both
non-cryptographic and cryptographic techniques. Non-cryptographic methods
transform the probe before it reaches the server, e.g., by discarding or
randomly selecting frequency components of the image~\cite{wang2022aaai,mi2022duetface,mi2023partialface}.
They add negligible overhead, but their protection is empirical, as it is
assessed against specific reconstruction attacks, and the server still obtains
the feature in the clear. Here we focus only on cryptographic approaches, as they offer much stronger security guarantees. The following overview classifies such approaches according to whether secure feature extraction is guaranteed or not.
Table~\ref{tab:related} compares GhostFaceMPC with existing schemes.  

\subsection{Secure matching over pre-computed features}
Most cryptographic PPFR systems let the client extract the probe feature in the
clear and protect only the matching. SCiFI~\cite{osadchy2010scifi} encodes faces as
patch-based binary vectors and evaluates a Hamming-distance test via additively homomorphic encryption and oblivious transfer, so that the client only learns whether a match occurs. For deep embeddings, Boddeti~\cite{boddeti2018secure} matches fixed-length features using FHE. HERS~\cite{engelsma2022hers} scales this approach to galleries of $10^8$~entries by compressing the representation, and
Blind-Match~\cite{choi2024blindmatch}
optimizes HE cosine similarity for 1:$N$ identification.
CryptoMask~\cite{cryptomask} combines HE and MPC to only reveal
whether the query matches an enrolled record. Funshade~\cite{ibarrondo2023funshade} uses function secret sharing to let two servers evaluate a thresholded distance in a single online round (a trusted dealer for
preprocessing is required, though).

Leaving the extractor outside the protected computation forces the client to
hold and run the network. This excludes thin clients and requires distributing the model to every
client, which is undesirable when the model is proprietary. 
In GhostFaceMPC the client only secret-shares the aligned image, never
sees the network, and receives a single decision bit. As
in~\cite{osadchy2010scifi,cryptomask,ibarrondo2023funshade}, no similarity score is released. 

\subsection{Secure feature extraction}
The earliest PPFR protocols~\cite{erkin2009privacy,sadeghi2009efficient}
already took encrypted images
as input. They computed the Eigenfaces projection under additively homomorphic
encryption and, in~\cite{sadeghi2009efficient}, the subsequent minimum search using garbled
circuits. Their linear extractor, however, is far less accurate than modern
CNNs. 
To our knowledge CryptoFace~\cite{cryptoface} is the only prior system that performs
deep feature extraction, matching and thresholding entirely on protected data.
 It uses FHE on a single server and, therefore, no non-collusion assumption is needed. To limit multiplicative depth, however, it
replaces the backbone with a mixture of shallow patch-based CNNs, and it reports about 20 minutes per image on CPU using an Amazon AWS \texttt{r5.24xlarge} machine equipped with 277 GB of RAM. GhostFaceMPC makes the complementary trade-off:
it assumes an honest majority among three non-colluding servers, and in
exchange it evaluates a standard lightweight face backbone with ArcFace head
in $2.43$\,s (Co-Host), $3.82$\,s (LAN) and $87.93$\,s (WAN), within
$0.32$ percentage points of the floating-point model on every benchmark, and it
additionally offers security with abort against a malicious server. We stress
that the latency figures were measured on different hardware and are
indicative of the order of magnitude only.

\begin{table*}[t]
\centering

\caption{Comparison of privacy-preserving face recognition systems.
``Parties'': entities taking part in the protected computation.
``Interaction'': whether the evaluation requires rounds of communication
between the parties beyond sending the input and receiving the output.
``Secure Extraction'': feature extraction is part of the protected computation.
``Verification'': probe matched against a claimed identity (1:1).
``Identification'': probe matched against the whole gallery (1:$N$).
}

\label{tab:related}
\begin{tabular}{llllcccc}
\toprule
System & Primitive & Parties & Interaction & Secure Extraction & Verification & Identification \\
\midrule
Erkin et al.~\cite{erkin2009privacy} & HE & client + server & \checkmark & linear (Eigenfaces) & & \checkmark \\
Sadeghi et al.~\cite{sadeghi2009efficient} & HE + GC & client + server & \checkmark & linear (Eigenfaces) & & \checkmark \\
Boddeti~\cite{boddeti2018secure}  & FHE & client + server & \checkmark & \\
HERS~\cite{engelsma2022hers} & FHE & client + server & & & \checkmark \\
CryptoMask~\cite{cryptomask} & HE + MPC & 1 server & \checkmark & & \\
Funshade~\cite{ibarrondo2023funshade} & FSS & 2 servers & \checkmark & & \checkmark & \checkmark \\
Blind-Match~\cite{choi2024blindmatch} & HE & 1 server & & & & \checkmark \\
CryptoFace~\cite{cryptoface} & FHE & 1 server & & \checkmark & \checkmark & \checkmark \\
\midrule
GhostFaceMPC & MPC & 3 servers & \checkmark & \checkmark & \checkmark & \checkmark \\
\bottomrule
\end{tabular}
\end{table*}

\section{Preliminaries}
\label{sec:preliminaries}
\subsection{Notation.}
The symbols $\mathbb{Z}$, $\mathbb{Z}^+$, and $\mathbb{R}$ denote the set of integers, positive integers, and real numbers, respectively. We represent the set $\{1,2,\dots,t\}$ by $[t]$ for $t \in \mathbb{Z}^+$. The ring of $k$-bit integers is represented by $\mathbb{Z}_{2^k}$. Given a real number $r$, $\lfloor r \rceil$ denotes rounding to the nearest integer. For two vectors $u,v$ of the same dimension $d$, $\langle u, v \rangle = \sum_{j \in [d]} u_j v_j$ denotes their inner product and $||u||$ the Euclidean norm of $u$.

\subsection{Secret Sharing.}
A secret-sharing scheme splits a secret $x$ into random ``shares'' that can be put together to reconstruct the secret, but individually do not reveal any information about the secret. An $(n,t)$-linear secret sharing scheme splits a secret $x$ (from some ring) into $n$ random shares $\{[[x]]_j\}_{j \in [n]}$ such that:
\begin{enumerate}
    \item Each share $[[x]]_j$ for $j \in [n]$ is an element of the ring.
    \item (Security) No set of $t$ or less shares reveals any information about $x$.
    \item (Correctness) Any set of $t+1$ distinct shares can be used to reconstruct the private input $x$. Formally, $\sum_{j \in I} c_j [[x]]_j = x$, where $I$ is an index set of size $t+1$, $c_j$'s are constants and addition is over the ring.
\end{enumerate}

Secret-sharing schemes such as Shamir secret-sharing~\cite{Shamir79} are extensively used in secure MPC protocols.

In this work, we adopt Replicated Secret Sharing (RSS), a variant in which each party holds a subset of multiple sub-shares rather than a single share. This replication simplifies reconstruction and enables efficient, non-interactive local operations.

\subsection{Secure Multiparty Computation.}
Secure Multiparty Computation (MPC)~\cite{Yao86, GMW87} is a class of protocols that allows $n > 1$ mutually distrusting parties to compute a public function $f$ on private inputs held by each party. Formally, consider $n$ parties $\{\mathcal{P}_i\}_{i \in [n]}$, each with a private input $x_i$ and an $n$-party MPC protocol secure against $t$ corruptions.
Then, to compute $f(x_1,\dots,x_n)$ (an ML model in our case), the parties can run the MPC protocol for this function with the security
guarantee that no subset of $t$ parties gets any information about
the data of other parties except what is revealed from the output of $f$.

Our work employs MPC protocols that evaluate a function $f$ represented as a circuit of gates (e.g., comparison, XOR, scalar addition, and multiplication). Each gate processes inputs from its input wires and produces the result on an output wire. The protocol maintains a key invariant: starting with linear secret-shares of the gate inputs, the participating parties interact via a secure protocol to obtain linear secret-shares of the output.

Security in these protocols is defined under two primary adversarial models:

\begin{itemize}
    \item Honest-but-curious (HbC): corrupted parties strictly follow the protocol but attempt to infer private inputs from internal states and transcripts.
    \item Malicious (M): corrupted parties may arbitrarily deviate from the protocol to compromise privacy or alter the result.
\end{itemize}

\subsection{Fixed-point Arithmetic.}
Because finite-bitwidth computers use a fixed bit allocation to represent real numbers, non-integers are inherently bounded in precision. As a result, real numbers with infinite representations (e.g., irrationals) cannot be exactly encoded on such hardware.

Fixed-point representation allows one to represent real numbers as integers from a ring by scaling the real number by a given scaling factor.
A real number $r \in \mathbb{R}$ can be represented as a \emph{fixed-point number} $x \in \mathbb{Z}_{2^k}$ with bitwidth $k$ and precision $f$, i.e., $f$ fractional bits and scale $2^f$, where
\[
    x = \lfloor(2^f \cdot r)\rceil \mod 2^k.
\]

The real value corresponding to a fixed-point
number $x$ with bitwidth $k$ and precision $f$ is $\frac{x}{2^f}$.

\section{Architecture and Threat Model}
\label{sec:arch-and-threat}
\subsection{Face Recognition}
Face verification consists of two stages: \emph{enrollment} and \emph{verification}. In the enrollment stage, the authentication server receives a reference face image and user identity from the client, computes its feature representation through a trained network, and stores the resulting feature-identity pair. Upon verification, the client sends a probe image and a claimed identity. The authentication server extracts the probe's feature representation via the same network and evaluates its match against the indexed reference feature.

In this work, we focus exclusively on the verification stage. Specifically, we assume that a trusted provider carries out the enrollment process and pre-computes the feature representations for all registered identities (see Section~\ref{subsec:arch} for further details).
Discriminative features are extracted using a CNN.
Given an input face image $x \in \mathbb{R}^{H \times W \times C}$, and a neural network $f_\theta(\cdot)$ parameterized by trainable weights $\theta$, the extracted feature vector is $y = f_\theta(x) \in \mathbb{R}^d$. To verify whether two face images $x_1,x_2$ belong to the same identity, we measure the similarity between their feature representations $y_1 = f_\theta(x_1)$ and $y_2 = f_\theta(x_2)$ as the cosine of the angle they span:
\[
    \mathsf{Score}(y_1,y_2) = \cos(y_1,y_2) = \frac{\langle y_1, y_2 \rangle}{||y_1|| \cdot ||y_2||}.
\]

The two images are classified as belonging to the same individual if $\mathsf{Score}(y_1,y_2) \ge T$, where $T$ is a predefined threshold. This decision process is formalized via the matching function:
\[
    \mathsf{Match}(y_1,y_2) =
    \begin{cases}
        1 & \text{if } \mathsf{Score}(y_1,y_2) \ge T,\\
        0 & \text{otherwise}
    \end{cases}
\]

This rule is equivalent to the squared-distance formulation frequently used in the literature: on the normalized features $u = y_1/||y_1||$ and $v = y_2/||y_2||$ one has $||u - v||^2 = 2 - 2\cos(y_1,y_2)$, so comparing the distance against a threshold $T_d$ is the same test as comparing the cosine against $T = 1 - T_d/2$, with the direction of the inequality reversed.

\subsection{Architecture}
\label{subsec:arch}
\begin{figure}[t]
    \centering
    \includegraphics[width=\columnwidth]{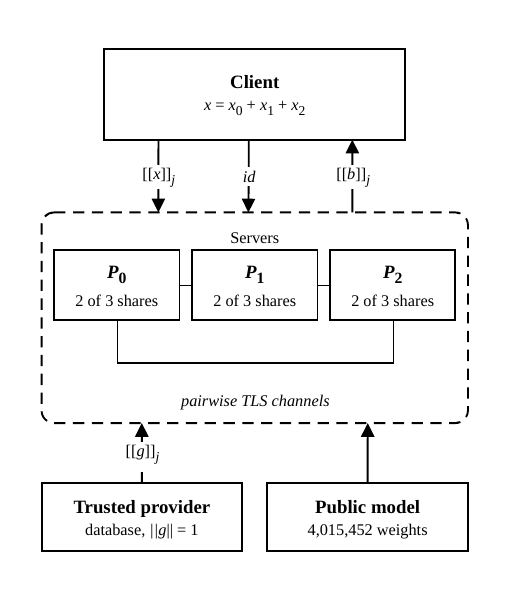}
    \caption{GhostFaceMPC architecture.}
    \label{fig:architecture}
\end{figure}

Our architecture is depicted in Figure~\ref{fig:architecture}. It involves five parties: a \emph{client} $\mathcal{C}$, three \emph{servers} $\mathcal{P}_0,\mathcal{P}_1,\mathcal{P}_2$, and a \emph{trusted provider} $\mathcal{TP}$ (which is active only at enrollment time). The three servers are connected by pairwise TLS channels and each hold the model, i.e., the $4{,}015{,}452$ convolution weights of $f_\theta(\cdot)$, in the clear. In the following, we call such values \emph{public}: they are known to the three servers, are not secret-shared, and are never sent to the client.
Secret values are held in replicated form: $x = x_0 + x_1 + x_2 \in \mathbb{Z}_{2^k}$ and $\mathcal{P}_i$ holds the pair $[[x]]_i = (x_{i-1}, x_i)$, indices modulo~$3$, so each component $x_j$ is held by $\mathcal{P}_j$ and $\mathcal{P}_{j+1}$. Any two components are uniform and independent of $x$, so a single server learns nothing about it, and when at most one server is corrupted every component has an honest holder.

\paragraph{Enrollment} The trusted provider computes the feature vector of every registered identity, stores it \emph{L2-normalized}, so that $||g|| = 1$ for each enrolled row $g$, and encodes it as a fixed-point vector $g = g_0 + g_1 + g_2$. It sends each server $\mathcal{P}_i$ its pair $[[g]]_i$. This happens once, before any request exists, after which $\mathcal{TP}$ may go offline: no single server ever holds an enrolled feature in the clear, and the database costs $N \times d \times 6 \times \lceil k/8 \rceil$ bytes of one-off upload for $N$ identities, outside every request. Normalizing at enrollment time removes the only square root of the decision rule from the secure computation, since $\cos(y,g) = \langle y, g \rangle / ||y||$ whenever $||g|| = 1$.

\paragraph{Request} The client holds the probe image and performs face detection and alignment locally. Then, it encodes the aligned $112 \times 112 \times 3$ RGB tensor as a fixed-point tensor $x$ over $\mathbb{Z}_{2^k}$ and splits it additively into $x = x_0 + x_1 + x_2$. In both profiles, it sends the pair $[[x]]_i$ to each server so that every component reaches both of its holders. The client therefore never transmits an image, and what a server receives is uniformly distributed and independent of the face. Alongside the shares, the client sends the identity it claims; that identity, and hence the database row index, is known by the servers in cleartext. Disclosing the claimed identity may be regarded as a privacy limitation, since it reveals to the servers which enrolled user is being authenticated. This leakage is inherent to 1:1 verification. When it is not acceptable, the identification mode (Section~\ref{sec:experiments}) compares the probe against all enrolled templates, so no identity is disclosed to the servers. The price is a matching cost that grows with the gallery size $N$ while the feature extraction cost remains the same.

\paragraph{Evaluation} The servers evaluate the network on the received replicated sharing $[[x]]$, obtaining a sharing $[[y]]$ of the probe feature $y = f_\theta(x)$. No server observes the input tensor, any intermediate activation, or the resulting feature. They then read the declared row $g_{\mathrm{id}}$ and compute the two inner products
\[
    s = \langle y, g_{\mathrm{id}} \rangle, \qquad n = \langle y, y \rangle,
\]
neither of which is opened. Because the row index is public, no oblivious database access is required and a request touches exactly one row, so its cost does not depend on the number $N$ of enrolled identities. From $s$ and $n$ the servers evaluate $\mathsf{Match}$ directly: since $\cos(y,g_{\mathrm{id}}) = s/\sqrt{n}$ and $T$ is public, for $T \ge 0$ the decision $\cos \ge T$ is equivalent to
\[
    s \ge 0 \;\wedge\; s^2 \ge C \cdot n, \qquad C = T^2,
\]
and for a negative one to $s \ge 0$ or $s^2 \le C \cdot n$. Squaring removes the square root, and the public constant $C$ multiplies a shared value locally, so no division is ever computed under MPC.

\paragraph{Response} The servers return the shares of $b = \mathsf{Match}(y, g_{\mathrm{id}})$: one component each in the semi-honest profile; in the malicious profile, only after verifying all pending multiplications, the pair $[[b]]_i$ each, so the client rejects the answer if two copies of a component differ. Since $[[b]]$ is the output of a multiplication, it is a fresh sharing and the client learns one bit per request. 

\subsection{Threat Model}
The protocol runs among three computing servers under an honest majority: at most one of the three servers may be corrupted, and the two remaining parties are assumed not to collude with it.

\paragraph{Corruption model} Corruption is \emph{static}: before the execution starts, the adversary corrupts at most one server and cannot move afterwards. 
We assume a HbC client $\mathcal{C}$, which follows the protocol but may try to infer information from what it receives. This is the standard assumption in prior privacy-preserving face recognition
work~\cite{erkin2009privacy,sadeghi2009efficient,osadchy2010scifi,boddeti2018secure,engelsma2022hers,cryptomask,ibarrondo2023funshade,cryptoface}. The view of $\mathcal{C}$ is its own input plus a fresh sharing of the decision bit, so $\mathcal{C}$ learns $b$ and nothing else. We do not consider a malicious client that submits malformed inputs. 
The trusted provider $\mathcal{TP}$ is honest and acts only during enrollment, after which it may go offline. The three servers communicate over pairwise authenticated and confidential channels. Non-collusion among the servers is the assumption the whole construction is bought with: any two of them jointly reconstruct every sharing, and hence the probe, the feature and the database.

\paragraph{Honest-but-curious profile} In the HbC profile the corrupted server follows the protocol and tries to infer secrets from its internal state and from the transcript. Its view consists of its own share components and of the messages it receives, all of which are uniformly distributed conditioned on the public data, and can therefore be simulated without any knowledge of the secrets: the adversary learns nothing about the probe image, about any intermediate activation, about the probe feature, about the enrolled rows, or about the decision bit itself, which is never reconstructed on the servers' side.

\paragraph{Malicious profile} The corrupted server is \emph{active}: it may deviate arbitrarily from the protocol, altering its inputs, messages or response, or halting at any point. We require \emph{security with abort} at $40$-bit statistical security. Privacy is as in the HbC profile: even by deviating, the adversary learns nothing about the probe, the features, the enrolled rows or the decision, and whether a request aborts does not depend on any of them. Except with probability $2^{-40}$, any deviation is detected and the request aborts. Hence, whenever the client receives a decision, it is the correct one, computed on the probe the client sent and on the enrolled row of the identity it declared: a corrupted server can neither replace these inputs nor alter the result. It can, however, always force an abort; we therefore don't guarantee output delivery.

\section{GhostFaceMPC}
\label{sec:ghostfacempc}
In this section, we present GhostFaceMPC, detailing its model architecture, implementation details, and optimizations.

\subsection{Implementation.}
\label{subsec:implementation}
GhostFaceMPC is implemented using MP-SPDZ~\cite{mp-spdz}, a C++ framework for multi-party computation (MPC) supporting various protocols and security models.

In this work, we employ the \emph{replicated-ring} protocol, a three-party MPC scheme secure in the HbC setting with an honest majority. This protocol relies on the replicated secret sharing scheme of Araki et al.~\cite{araki}, extended to $\mathbb{Z}_{2^k}$ ring domains as formalized in ABY3~\cite{ABY3}, with fixed-point rounding optimizations from Dalskov et al.~\cite{dalskov2020}. 

To achieve malicious security with abort under honest majority,  we employ the \emph{ps-rep-ring} protocol, which makes use of the post-sacrifice verification technique introduced in \cite{FLNW17}.

\subsection{Model.}
\label{subsec:model}
We adopted GhostFaceNet~\cite{ghostfacenet}, trained on the MS1MV3 dataset~\cite{lfr2019,msceleb1m}, as the baseline for our model. Specifically, we use the GhostFaceNetV1-1 variant (width $1.3$, stride $1$) with an ArcFace head~\cite{arcface}, which maps an aligned $112 \times 112 \times 3$ input to a $512$-dimensional embedding. For a detailed description of the model architecture, we refer to the original paper.

In our architecture the model parameters are public (i.e., known in the clear to the servers), and only the input tensor and the values computed from it (activations and embedding) are secret-shared. Every linear layer therefore reduces to a secret-by-public inner product, which is computed locally and requires a single rounding per output rather than a multiplication triple per multiply-accumulate.
Under this setup the residual communication cost is dominated by the ring itself: a party re-shares $\lceil k/8 \rceil$ bytes per exchanged element, so the width $k$ of $\mathbb{Z}_{2^k}$ multiplies the whole cost of an inference, and the fixed-point precision $f$ additionally sets the declared bitlength of each rounding.
Our starting configuration, a $64$-bit ring with $f = 16$, is therefore twice as expensive per element as it needs to be, and narrowing the ring is the single largest bandwidth lever available before any protocol-level optimization. This is the motivation for quantizing the model.
Naive post-training quantization of the float teacher to a $32$-bit ring with $f = 6$ does not survive this transition: the resulting model loses more than $30$ percentage points of verification accuracy. The cause is the accumulation of rounding error along the operations of the network. Recovering the accuracy therefore requires training the model on the exact arithmetic that the servers will execute, rather than on a floating-point idealization of it.

Our resulting model is obtained by knowledge distillation. The \emph{float} teacher is the pretrained GhostFaceNetV1-1 checkpoint (MS1MV3, ArcFace). The student starts from the teacher's weights and is fine-tuned through a differentiable forward pass that reproduces the integer ring arithmetic of the MPC program. Training uses the same MS1MV3 training dataset. The loss combines the cosine distance and the mean squared error between teacher and student embeddings, a feature-matching term at intermediate layers, and a small penalty that keeps the deferred accumulators within their overflow bounds.

We used the Adam optimizer with learning rate $5 \times 10^{-5}$, warm-up followed by cosine decay, batch size 128. The checkpoint is selected by the mean teacher-student embedding distance on 512 held-out faces. The quantized model and the full quantization code, including the exact loss terms and hyperparameters, are available in the repository.

\paragraph{Fixed-point precisions} We used two fixed-point encodings. Every secret value (the input tensor, each intermediate activation and the embedding) is encoded with $f$ fractional bits, i.e. on the grid $2^{-f}$ with scale $S = 2^f$; we call $f$ the \emph{activation precision}. The public convolution and depthwise kernels, after batch-normalization folding, have their own number of fractional bits $f_w$ and sit on the finer grid $2^{-f_w}$ with scale $S_w = 2^{f_w}$; we call $f_w$ the \emph{weight precision}. Decoupling the two lets the kernels keep a resolution the activations do not need, and because the kernels are never shared, this costs no communication. A secret-by-public product therefore lands on the scale $S S_w = 2^{f + f_w}$, and the single rounding of each linear output shifts it right by $f_w$ bits, back onto the activation grid.
The resulting profile is a $32$-bit ring with activation precision $f = 6$, weight precision $f_w = 9$ and declared rounding length $r = 15$, with nearest rounding throughout. Its accuracy is reported in Table~\ref{table:summary} and discussed in Section~\ref{sec:experiments}.

\subsection{General Techniques and Optimizations.}
\label{subsec:optimizations}
Under the setting described in Section~\ref{subsec:model}, the cost of an inference is not its arithmetic but its interaction. Every product of a secret activation by a public coefficient is local, so most of what the three servers exchange is due to the fixed-point roundings that bring an accumulator back onto the activation grid and to the sign tests that the activations require; the few secret-by-secret products (squeeze-and-excitation gates, PReLU slope selection, matching) contribute the rest. In our first experiments, those two operations account for $93\%$ of the bytes of an inference, $74\%$ for rounding and $19\%$ for sign extraction.

Communication is the sum of the bytes sent by the three parties for one inference, a round is one online communication round of the compiled program, and the two sides of a comparison are measured in the same session on the same host, in alternating order, with three pools of five discarded warm-up requests followed by thirty measured requests per pool. Bytes and rounds are deterministic properties of the compiled circuit and are therefore reported as exact figures rather than as medians; wall and CPU times are reported as the median over the measured requests.

\paragraph{Public parameters and one rounding per linear output}
The model is not secret to the servers: every kernel, bias, normalization coefficient and PReLU slope is known in the clear to all three servers.
Within the network, only the image is secret, and with it every activation computed from it. This matters because in MPC the cost of a multiplication depends on what is being multiplied.

To multiply a secret value by a public constant, each server multiplies its own share by that constant; the results are again valid shares of the product, and no message is exchanged. To multiply two secret values, the servers have to exchange messages and consume precomputed randomness, which costs both bytes and rounds. A convolution or a dense layer consists entirely of products of the form \emph{activation}~$\times$~\emph{weight}. Since, in our architecture, every weight is public, all of these products are of the first, communication-free kind, and each server computes the whole layer on its own shares, locally, as it would in the clear. Secret-by-secret products occur only outside the linear layers, in the squeeze-and-excitation gates, in the selection of the PReLU slope and in the matching, and they are few compared with the multiply-accumulate operations of the network.

Within a linear layer, the only step that still requires communication is the rescaling that fixed-point arithmetic needs after a product (described later in this section), and it is applied once to each output value, not once per multiplication: a $K\times K$ convolution over $C_{\mathrm{in}}$ input channels performs $C_{\mathrm{in}}K^2$ multiply-accumulate operations for every output, and communicates only for that single rescaling. 
The communication of a linear layer therefore grows with its number of outputs, not with its number of multiply-accumulate operations. The multiplications themselves still have to be computed, but since they involve no interaction they are ordinary integer arithmetic on each server, and can be vectorized, tiled and spread across threads. Every optimization described in the rest of this section builds on this property.

\paragraph{Vectorization}
Since the weights are public, every convolution is local, but a direct implementation still evaluates it one scalar product at a time, with a separate accumulator and separate instructions for each output channel. In our first implementation this local arithmetic, rather than communication, dominated the wall time.
The same layout also batches the communication. A direct implementation issues some interactive operations one element at a time inside a loop, such as the per-channel rescaling of global average pooling, and each element then waits for its own sequence of rounds. We instead issue all the roundings and sign tests of a layer as a single vector operation. Their messages travel together, so a layer costs the rounds of one rounding regardless of how many outputs it has. Batching removes $95.6\%$ of the rounds (Table~\ref{table:ablation}).


\paragraph{Batch-Normalization folding}
The batch normalization layer for a channel $c$ is described by:
\[
    y_c = \gamma_c \cdot \frac{x_c - \mu_c}{\sqrt{\sigma_c^2 + \varepsilon}} + \beta_c
\]
where $x_c$ is the input, $\mu_c,\sigma_c$ are constants and, $\gamma_c,\beta_c$ are the parameters learned during training.

Since all terms except the input are fixed at inference time, this layer can be written as an affine map $y_c = a_cx_c + b_c$, where $a_c = \frac{\gamma_c}{\sqrt{\sigma_c^2 + \varepsilon}}$ and $b_c = \beta_c - a_c\mu_c$.
The coefficients $a_c,b_c$ are public, so multiplication does not require communication. What it costs is a change of scale. The input $x_c$ is represented by an integer $\lfloor x_c 2^f\rceil$, shared among the three servers. On the other hand, the coefficient $a_c$ is represented by the integer $\lfloor a_c 2^{f_w} \rceil$, so their product carries the correct value on the wrong grid, at scale $2^{f+f_w}$. For this reason, we need to shift it back by $f_w$ bits, before any subsequent layer can consume it.
This shift is a rounding and, as such, requires communication between the parties.

In our architecture, that normalization appears always after a convolution layer, which is itself an affine map with public coefficients. Since a composition of affine maps is again affine, the two layers can be merged into a single convolution of weights and bias:
\[
    W'_{c} = a_c W_{c}, \qquad b'_c = (b_c - \mu_c)a_c + \beta_c,
\]
whose one rounding now covers both. The merge is carried out ahead of time, while the circuit is compiled, and only the numerical values of the public weights change: the network evaluated by the servers has one layer fewer and one rounding fewer per element, and the result is the same. The merge requires that the convolution output feed nothing but the normalization, since otherwise its unnormalized values are still needed elsewhere in the network. Of the $21$ batch normalizations that are evaluated as separate layers in the naive circuit, $20$ qualify; in the remaining case the convolution output is also read by a second branch, and the normalization stays a layer of its own. 

\paragraph{Bit extraction} Rounding and sign extraction, the two non-linear operations in the circuit, cannot be evaluated locally on arithmetic shares. Rounding requires recovering the carry bits generated when the three shares are summed, while sign extraction requires obtaining the most significant bit of that sum. Because bit extraction on a secret-shared value $x = x_0 + x_1 + x_2$ is non-linear with respect to its shares, no party can compute its portion locally. 
To solve this problem, we perform bit extraction over \emph{binary} secret sharing, where a value is reconstructed as the XOR $b = b_0 \oplus b_1 \oplus b_2$ of its shares and an adder is realized as a circuit of XOR and AND gates. To convert arithmetic shares to boolean shares, during the online phase, we employ the \emph{daBits} technique introduced by Rotaru~\emph{et al.}~\cite{Rotaru2019}. Using this technique requires the introduction of an offline phase in which the parties generate the daBits they need during the online phase.

\paragraph{Rounding fusion}
As previously mentioned, rounding is the most expensive operation. Reducing the number of rounding operations allows us to significantly decrease the bandwidth required by our protocol.

Many roundings exist only to bring a value back to the activation scale before it is consumed by an operation whose result will be rounded anyway. If such a value is left unreduced and carried forward, the later rounding can shift by the combined amount and perform both rescalings at once. The final result is the same up to rounding, and one interactive operation per element disappears. We apply this in three places:

\begin{enumerate}
    \item Convolution followed by PReLU: PReLU computes $y = x$ if $x \ge 0$ and $y = \alpha x$ otherwise, with a public slope $\alpha$. The sign of a number does not change when it is multiplied by a positive scale. The branch can therefore be decided directly on the convolution's unreduced accumulator. The accumulator is then multiplied by the selected slope, and a single rounding removes both the convolution's scale and the slope's scale. Two roundings become one.
    
    \item Squeeze-and-excitation followed by a convolution: a squeeze-and-excitation block multiplies a feature map by a secret per-channel gate. That product is fed to a $1 \times 1$ convolution with public weights, which must reduce its own output anyway. We keep the product unreduced, let the convolution accumulate it, and perform one rounding for both.
    
    \item Residual additions: addition commutes with scaling. The two branches of a residual connection can therefore be added while still unreduced and reduced once, instead of being reduced separately and then added.
\end{enumerate}

Unfortunately, an unrounded value could be risky. If its magnitude exceeds what the ring can represent, it wraps around modulo $2^k$ and the result becomes meaningless. A postponed rounding is therefore allowed only when we can show that the unreduced value still fits. For instance, a convolution and the PReLU that follows it are fused only if
\[
    e + f + f_w + b_\alpha \le k - 2
\]
where $e$ is the number of integer bits the accumulator can occupy and $b_\alpha$ is the number of bits of the quantized slope. How many pairs can be fused therefore depends on how tightly we can bound $e$.

Using a \emph{static} bound, we assume that every input $x_t$ of a layer reaches the bound $|x_t| \le 2^{r-f-1}$ that any value with $r$ total and $f$ fractional bits satisfies. For output channel $o$, with public weights $W_{t,o}$ over all inputs $t$ that reach it (every input channel and kernel position) and bias $b_o$, the triangle inequality gives
\[
    B_{\mathrm{static},o} = 2^{r-f-1} \sum_t |W_{t,o}| + |b_o|.
\]
Real inputs never come close to this case. The bound overestimates the actual activations so we apply a \emph{dynamic} bound to each layer by measurement instead.

For each linear layer $l$ we run the network on faces disjoint from every evaluation dataset and record its peak $P_l$, the largest magnitude of any pre-activation over all channels, positions and images. The calibrated bound multiplies that peak by $2^2 = 4$, a two-bit safety margin:

\[
    B_{\mathrm{cal},l} = \max\bigl(1, 4 P_l,\ \max_o |b_{l,o}|\bigr) .
\]
The second term keeps the bound above every bias $b_{l,o}$ of the layer, since a channel that is nearly inactive on the calibration faces can have a measured peak below its own bias. The third keeps $\lceil \log_2 B_{\mathrm{cal},l} \rceil$, the number of integer bits the bound declares, non-negative.

\paragraph{Two-summand rounding}
A rounding by $n$ bits cannot be applied to each share on its own. In replicated secret sharing, a value is $x = x_0 + x_1 + x_2$, and shifting the components separately loses the carries produced when they are added. The servers recover those carries with a binary circuit over the bits of the components and convert each carry bit back to the arithmetic domain with a bit injection. With three addends, a carry can be 0, 1, or 2, so each carry costs 2 injections.
The replicated structure allows a cheaper split, which has been used for local share conversion~\cite{ABY3} and for probabilistic truncation~\cite{dalskov2020}; here we use it for exact rounding.
Party $\mathcal{P}_0$ holds $x_0$ and $x_2$, so it alone knows $A = x_0 + x_2$, while $\mathcal{P}_1$ and $\mathcal{P}_2$ both hold $B = x_1$. With two addends every carry is 0 or 1, so it takes a single injection. The arithmetic part shifts $A$ and $B$ separately. Both holders of $B$ know $\lfloor B/2^n \rfloor$ and write it locally. Only $\mathcal{P}_0$ knows $\lfloor A/2^n \rfloor$, and it re-shares that value to $\mathcal{P}_1$, masked with randomness it shares with $\mathcal{P}_2$, at the cost of one ring element and one round. Nothing is opened, and $\mathcal{P}_1$ sees a uniformly random element.

\section{Experiments}
\label{sec:experiments}
In this section, we present our experimental results.
Following recent experimental setups in the MPC literature (e.g.,~\cite{MD-ML, morita2025maestro}), we evaluated our protocols under two network environments, emulated with the Linux traffic control tool \texttt{tc}: a LAN setting with 1 Gbps bandwidth and 0.1 ms RTT, and a WAN setting with 100 Mbps bandwidth and 100 ms RTT. We additionally report a Fast-WAN setting with 1 Gbps bandwidth and 10 ms RTT, and a Co-Host setting, in which the three parties run on the same machine without any traffic shaping.

All experiments were executed on a single machine: a $32$-core virtual machine with $24$ GB of RAM, backed by four Intel Xeon Gold 6238R sockets at $2.20$ GHz.

\subsection{Test Datasets.}
Following AdaFace~\cite{AdaFace}, we evaluate the quantized model and baseline methods across five standard benchmark datasets: LFW~\cite{LFW}, AgeDB~\cite{AgeDB}, CALFW~\cite{CALFW}, CPLFW~\cite{CPLFW}, and CFP-FP~\cite{CFP-FP}. All images are processed at a resolution of $112 \times 112$. The CFP-FP dataset contains 7,000 image pairs (totaling 14,000 face images), whereas the remaining datasets consist of 6,000 pairs (12,000 face images) each.

\subsection{Baselines.}
To the best of our knowledge, there is no prior face-recognition system that evaluates a deep CNN extractor and matching under MPC. As a result, we adopt a naive implementation in MP-SPDZ~\cite{mp-spdz} and EzPC~\cite{EzPC} (using \emph{Porthos}~\cite{cryptflow}) as our baselines, both running the same checkpoint on the same secret input (Table~\ref{table:summary}). The naive MP-SPDZ baseline is a straightforward secret-shared realization of the network, scalar kernels, one rounding per linear output, no fusion and no graph rewriting, over a $64$-bit ring with $f = 20$. The Porthos baseline is reached through the Athos ONNX frontend of EzPC with exact-rounding variant. A single-party TensorFlow float32 run is the plaintext floor.
 The baseline rows run at $64$-bit precision, which reproduces the decisions of the float model, and hence inherit the TensorFlow accuracy; the GhostFaceMPC rows are those of its quantized model, at most $0.32$ percentage points (CPLFW) below the float teacher.

\subsection{Optimizations Analysis}
Table~\ref{table:ablation} depicts the contribution of each optimization described in Section~\ref{subsec:optimizations}. The first row is the naive MP-SPDZ baseline of Table~\ref{table:summary}. Every following row adds one optimization to the row above it, and the last row is GhostFaceMPC. All rows run the same verification request under the HbC in a Co-Host environment, with $8$ threads per party. Bandwidth and rounds refer to the online phase. The daBits generated in the offline phase are reported in a separate column.
Vectorization removes $95.6\%$ of the rounds and $87\%$ of the wall time. Folding $20$ of the $21$ batch normalizations removes as many roundings per element of their outputs and saves a further $5.4\%$ of the bandwidth and $10\%$ of the rounds.
Ring reduction is the largest single saving. It cuts the online bandwidth by $54\%$ and the offline daBits by half, and it accounts for $63\%$ of the bandwidth saved between the first and the last row. Rounds fall by $13\%$ as well, because the carry circuits of a rounding become shorter on the narrower declared widths. The three fusions together save $14\%$ of the bandwidth, $19\%$ of the rounds and $28\%$ of the wall time. At this profile the bound of Section~\ref{subsec:optimizations} admits $32$ of the $41$ Conv--PReLU pairs, all $7$ squeeze-and-excitation pairs and $5$ of the $16$ residual additions. The squeeze-and-excitation fusion saves the most time ($17\%$), since the rounding it removes acts on full feature maps.
Two-summand rounding needs one daBit per carry instead of two. It therefore saves $8\%$ of the online bandwidth and $44\%$ of the offline daBits. It also adds one round to each rounding, for the re-sharing by $\mathcal{P}_0$, which raises the rounds by $19\%$. In the Co-Host setting the bandwidth saving prevails and the wall time still drops by $8.0\%$.
Overall, the optimizations reduce the online bandwidth by $70.3\%$, the rounds by $96.7\%$ and the wall time by $94.6\%$ relative to the naive baseline.

\begin{table*}
\centering
\caption{Optimizations analysis: one verification request, HbC, Co-Host. Each row adds one optimization to the row above it; $\Delta$ is the change relative to the row above, except for the GhostFaceMPC row, where it is relative to the naive MP-SPDZ baseline.}
\label{table:ablation}

\begin{tabular}{l l c r r r r r r r}
\toprule
 & & & \multicolumn{6}{c}{\textbf{Online phase}} & \textbf{Offline phase} \\
\cmidrule(lr){4-9} \cmidrule(lr){10-10}
\textbf{Step} & \textbf{Ring} & \textbf{Layers} & \textbf{Rounds} & $\boldsymbol{\Delta}$ & \textbf{Bandwidth (MB)} & $\boldsymbol{\Delta}$ & \textbf{Time (s)} & $\boldsymbol{\Delta}$ & \textbf{Bandwidth (MB)} \\
\midrule
MP-SPDZ (naive)               & 64 & 236 & 42784 & --         & 1052.54 & --         & 45.00 & --         & 920.04 \\
+ Vectorization               & 64 & 236 & 1873  & $-95.6\%$  & 913.40  & $-13.2\%$  & 6.06  & $-86.5\%$  & 919.92 \\
+ BN folding                  & 64 & 216 & 1693  & $-9.6\%$   & 864.11  & $-5.4\%$   & 4.90  & $-19.2\%$  & 869.50 \\
+ Ring reduction              & 32 & 216 & 1480  & $-12.6\%$  & 395.91  & $-54.2\%$  & 3.66  & $-25.3\%$  & 434.75 \\
+ Conv--PReLU fusion    & 32 & 216 & 1296  & $-12.4\%$  & 370.76  & $-6.4\%$   & 3.27  & $-10.5\%$  & 334.65 \\
+ SE--Conv fusion             & 32 & 216 & 1240  & $-4.3\%$   & 342.35  & $-7.7\%$   & 2.70  & $-17.5\%$  & 304.69 \\
+ Residual-Add fusion         & 32 & 216 & 1206  & $-2.7\%$   & 339.64  & $-0.8\%$   & 2.65  & $-2.1\%$   & 299.90 \\
+ Two-summand rounding        & 32 & 216 & 1433  & $+18.8\%$  & 312.35  & $-8.0\%$   & 2.43  & $-8.0\%$   & 168.76 \\
\midrule
\textbf{GhostFaceMPC}         & 32 & 216 & 1433  & $-96.7\%$  & 312.35  & $-70.3\%$  & 2.43  & $-94.6\%$  & 168.76 \\
\bottomrule
\end{tabular}
\end{table*}

\subsection{Face Identification.}
In this paper, we focus mainly on the face verification task, as discussed in Section~\ref{sec:arch-and-threat}. However, GhostFaceMPC can be readily extended to face identification scenarios, such as one-to-many face matching. In Table~\ref{table:identification-benches}, we report the experimental results for a $1 : 128$ closed-set rank retrieval task under MPC, using a randomly selected subset of LFW consisting of 128 pairs. The cost is close to that of verification (Table~\ref{table:verification-benches}): in both cases the dominant cost is the feature extraction of the probe image, which is performed only once, while identification merely evaluates $N$ matching functions instead of one, an overhead linear in the gallery size $N$.

\begin{table*}
\centering
\caption{Comparison between GhostFaceMPC and baselines.}
\label{table:summary}

\begin{tabular}{l l c c c c c c c c c c}
\toprule
\multirow{2}{*}{\textbf{Method}} & \multicolumn{2}{c}{\textbf{Settings}} & \multicolumn{6}{c}{\textbf{Dataset}} & \multirow{2}{*}{\textbf{Rounds}} & \multirow{2}{*}{\textbf{Time (s)}} & \multirow{2}{*}{\textbf{Bandwidth}} \\
\cmidrule(r){2-3} \cmidrule(lr){4-9}
 & Security & Ring & LFW & AgeDB & CALFW & CPLFW & CFP-FP & Avg & & \\
\midrule
\textbf{Tensorflow}
 & None & -- & 99.72 & 97.80 & 95.80 & 91.92 & 96.39 & 96.33 & -- & 0.16 & -- \\
\midrule
\textbf{EzPC} (Porthos)
 & HbC & 64 & 99.72 & 97.80 & 95.80 & 91.92 & 96.39 & 96.33 & 2152 & 258.96 & 17.97 GB \\
\midrule
\textbf{MP-SPDZ} (naive)
 & HbC & 64 & 99.72 & 97.80 & 95.80 & 91.92 & 96.39 & 96.33 & 42784 & 45.00 & 1.05 GB \\
 \midrule
\multirow{2}{*}{\textbf{GhostFaceMPC}} 
 & HbC  & 32 & 99.68 & 97.67 & 95.85 & 91.60 & 96.37 & 96.23 & 1433 & 2.43 & 312.35 MB \\
 & M  & 32 & 99.68 & 97.67 & 95.85 & 91.60 & 96.37 & 96.23 & 1391 & 5.53  & 883.27 MB \\
\bottomrule
\end{tabular}
\end{table*}

\begin{table*}
\centering
\caption{Face Verification Benchmarks: offline and online phases.}
\label{table:verification-benches}
\begin{tabular}{l l r r r r r r}
\toprule
\multirow{2}{*}{\textbf{Security}} & \multirow{2}{*}{\textbf{Setting}}
 & \multicolumn{2}{c}{\textbf{Rounds}}
 & \multicolumn{2}{c}{\textbf{Time (s)}}
 & \multicolumn{2}{c}{\textbf{Bandwidth (MB)}} \\
\cmidrule(lr){3-4} \cmidrule(lr){5-6} \cmidrule(lr){7-8}
 & & Offline & Online & Offline & Online & Offline & Online \\
\midrule
\multirow{4}{*}{\textbf{HbC}}
 & Co-Host  & 14  & 1433 & 0.90   & 2.43   & 168.76  & 312.35 \\
 & LAN      & 14  & 1433 & 1.49   & 3.82   & 168.76  & 312.35 \\
 & Fast-WAN & 14  & 1433 & 1.55   & 10.76  & 168.76  & 312.35 \\
 & WAN      & 14  & 1433 & 9.97   & 87.93  & 168.76  & 312.35 \\
\midrule
\multirow{4}{*}{\textbf{M}}
 & Co-Host  & 380 & 1391 & 28.41  & 5.53   & 5504.13 & 883.27 \\
 & LAN      & 380 & 1391 & 41.57  & 8.14   & 5504.13 & 883.27 \\
 & Fast-WAN & 380 & 1391 & 44.16  & 16.72  & 5504.13 & 883.27 \\
 & WAN      & 380 & 1391 & 244.07 & 124.68 & 5504.13 & 883.27 \\
\bottomrule
\end{tabular}
\end{table*}

\begin{table*}
\centering
\caption{Face Identification Benchmarks: offline and online phases.}
\label{table:identification-benches}
\begin{tabular}{l l r r r r r r}
\toprule
\multirow{2}{*}{\textbf{Security}} & \multirow{2}{*}{\textbf{Setting}}
 & \multicolumn{2}{c}{\textbf{Rounds}}
 & \multicolumn{2}{c}{\textbf{Time (s)}}
 & \multicolumn{2}{c}{\textbf{Bandwidth (MB)}} \\
\cmidrule(lr){3-4} \cmidrule(lr){5-6} \cmidrule(lr){7-8}
 & & Offline & Online & Offline & Online & Offline & Online \\
\midrule
\multirow{4}{*}{\textbf{HbC}}
 & Co-Host  & 14  & 1496 & 0.89   & 2.44   & 168.76  & 312.35 \\
 & LAN      & 14  & 1496 & 1.50   & 3.82   & 168.76  & 312.35 \\
 & Fast-WAN & 14  & 1496 & 1.55   & 11.06  & 168.76  & 312.35 \\
 & WAN      & 14  & 1496 & 9.98   & 90.83  & 168.76  & 312.35 \\
\midrule
\multirow{4}{*}{\textbf{M}}
 & Co-Host  & 380 & 1454 & 28.24  & 5.66   & 5504.14 & 888.54 \\
 & LAN      & 380 & 1454 & 42.61  & 8.30   & 5504.14 & 888.54 \\
 & Fast-WAN & 380 & 1454 & 44.05  & 17.13  & 5504.14 & 888.54 \\
 & WAN      & 380 & 1454 & 245.42 & 130.73 & 5504.14 & 888.54 \\
\bottomrule
\end{tabular}
\end{table*}

\section{Conclusion}
\label{sec:conclustion}
This paper introduced GhostFaceMPC, to our knowledge the first end-to-end
privacy-preserving face recognition system running both a deep CNN extractor
and matching under secure multi-party computation: the client secret-shares
the aligned face among three non-colluding servers and learns only the
decision. With a public model, secure inference cost is set by fixed-point
roundings, not multiplications, so performance requires co-designing model and
circuit: training on the servers' exact ring arithmetic and removing or
cheapening every unneeded rounding.
The main limitation is wide-area latency, dominated by about 1,400 rounds per
inference. Reducing the round complexity of truncation and comparison is the
key direction, alongside enrollment under MPC and large-gallery
identification. GhostFaceMPC shows that strong cryptographic protection of
biometric data is compatible with practical face recognition.

\bibliographystyle{IEEEtran}
\bibliography{references}

\newpage
\vspace{11pt}

\end{document}